\documentstyle[12pt,prc,aps,floats,epsfig]{revtex}
\tighten
\begin{document}
\draft
\title{Comparison between chiral and meson-theoretic\\
nucleon-nucleon potentials through (p,p$'$) reactions}
\author{F. Sammarruca and D. Alonso}
\address{Physics Department, University of Idaho, Moscow, ID 83844, U.S.A}
\author{E.J. Stephenson}
\address{Indiana University Cyclotron Facility, Bloomington, IN
47408 U.S.A.}
\date{\today}
\maketitle
\begin{abstract}
We use proton-nucleus reaction data at intermediate energies to test
the emerging new generation of chiral nucleon-nucleon (NN) potentials.
Predictions from a high quality one-boson-exchange (OBE) force are
used for comparison and evaluation.  Both the chiral and OBE models fit
NN phase shifts accurately, and the differences between the two
forces for proton-induced reactions are small.  A comparison to a
chiral model with a less accurate NN description sets the scale for
the ability of such models to work for nuclear reactions.
\end{abstract}
Chiral perturbation theory ($\chi$PT) offers a way to describe
phenomena at nuclear physics energies that is consistent with the
symmetries of the underlying theory of strong interactions (QCD). In this low
momentum regime, QCD itself is non-perturbative.  In $\chi$PT, one expands
chiral $\pi N$ Lagrangians in powers of the relevant
momenta or masses ({\it e.g.}, pion mass), relative to the QCD
scale at $\Lambda_{\rm QCD}\sim 1$~GeV. The nucleon-nucleon (NN) force
can then be derived from chiral Lagrangians by taking into account all the pion-exchange
diagrams which contribute to the NN interaction at a given chiral order.
NN potentials based on $\chi$PT are thus best suited for applications at
the lowest energies.
An important energy range opens near 200~MeV where the
impulse approximation treatment becomes a meaningful test of the
ability of an NN  potential to describe nucleon-induced reactions on
nuclei.  Only recently has a chiral NN potential
become available through the work of Entem and Machleidt (EM) \cite{EM}
that accurately reproduces NN phase shifts at
this energy.  In this paper, we present the first tests
of a chiral potential using (p,p$'$) reactions.

Proton-nucleus elastic and inelastic scattering
to selected transitions can be a
``laboratory'' for the evaluation of NN interactions \cite{S1,S2,S4}.
Through polarization observables, (p,p$'$) reactions are selectively
sensitive to specific amplitudes in the effective interaction
\cite{Moss,Blez} constructed by placing that NN interaction in the
nuclear medium.  Natural-parity transitions,
for example, sample the central and
spin-orbit terms in the isoscalar channel.  ``Spin-flip'' states can
examine all the spin-dependent terms of either isospin.  Here we
will use specific (p,p$'$) transitions from a study at 200~MeV
\cite{S1,S2} to compare two recent $\chi$PT NN potentials
\cite{EM,Eb1} with calculations based on a more
conventional one-boson-exchange (OBE) potential.  The two $\chi$PT
potentials differ in the precision with which they reproduce NN
phase shifts.  This will calibrate for us the quality of agreement
needed in $\chi$PT to describe reactions such as proton-nucleus
elastic and inelastic scattering.  Previous studies using
conventional potentials
demonstrated that good reproduction of (p,p$'$) observables
depends on a high quality representation of the NN data as well
\cite{S1}.

The $\chi$PT potential of EM \cite{EM} is based on
a heavy baryon expansion scheme, where nucleon fields are  represented by
2-component spinors.  This makes a relativistic treatment of nucleons in
nuclear matter (Dirac-Brueckner-Hatree-Fock, or DBHF, approach)  not
feasible. Thus we will only include
Brueckner-Hartree-Fock (BHF) medium effects when calculating density-dependent
effective interactions.

The $\chi$PT expansion  includes
 1-$\pi$ and 2-$\pi$ diagrams from effective chiral Lagrangians, with relativistic
corrections, through third order.  The short-range repulsion is
described by including contact terms up through fourth order.
To be suitable for iteration in a Lippmann-Schwinger equation,
the potential is regularized through a set of cutoff masses.
The 46
model parameters were adjusted to match the NN phase shift solution
from the Nijmegen group \cite{Nij}.  The resulting agreement is
excellent at energies below 300~MeV.

The OBE model that we have chosen for comparison is the
CD-Bonn potential \cite{Adv}.
Medium modifications arise from an effective nucleon mass
(produced in a self-consistent calculation of nuclear matter
saturation properties) and a spherically averaged Pauli
blocking operator
\cite{HT}.  The resulting
density-dependent G-matrix is transformed into a Yukawa
representation for use in distorted wave impulse approximation
(DWIA) (p,p$'$) reaction calculations (see App.\ A of \cite{S1}).

The DWIA calculations are made with the programs LEA \cite{LEA}
for natural parity transitions and DWBA86 \cite{DWBA86} for
un-natural parity.  The distortions are generated from the
density-dependent effective interaction using the folding model.
The formfactors are chosen to conform to (e,e$'$) measurements for
the same transitions.  The nuclear matter density is extracted
from the charge density by unfolding the formfactor of the proton
\cite{deVries}.  Additional details may be found in Refs.\
\cite{S1,S2}.

\begin{figure}[t]
\vspace{-9cm}
\hspace{-2.0cm}
\epsfig{figure=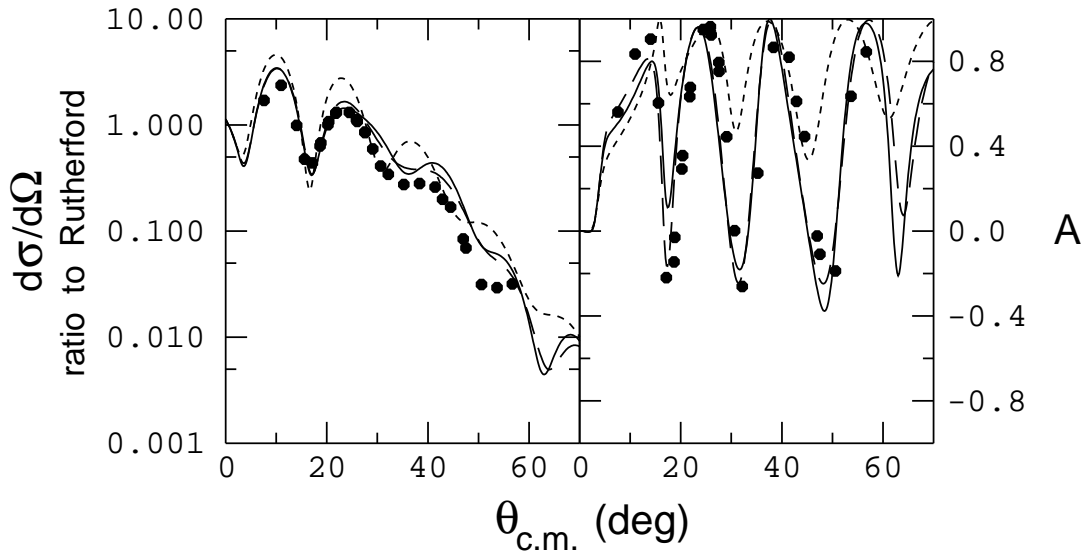,width=16cm}
\vspace{-4cm}
\caption{Measurements for proton elastic scattering cross section
(shown as the ratio to the Rutherford cross section) and analyzing
power from Ref.~[14].  Calculations are
based on the chiral NN potential of Ref.~[1] (solid curves) and the
CD-Bonn potential of Ref.~[9] (long-dashed curves).
Both calculations contain
BHF density dependence.  The short-dashed curves are
CD-Bonn potential
calculations with no density dependence included.}
\label{one}
\end{figure}

\begin{figure}
\vspace{-9cm}
\hspace{-2.0cm}
\epsfig{figure=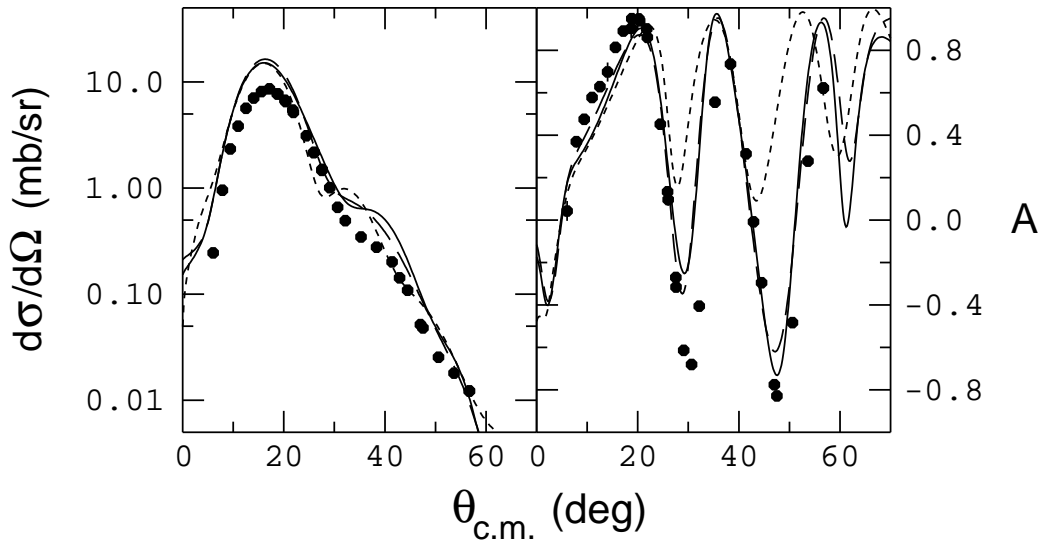,width=16cm}
\vspace{-4cm}
\caption{Measurements for proton inelastic scattering cross section
and analyzing power for the $3^-$ state in $^{40}$Ca at 3.736~MeV
from Ref.~[14].
The curves are the same as in Fig.~1.}
\label{two}
\end{figure}

Figures~\ref{one} and \ref{two} show density-dependent
calculations of the differential cross section and vector analyzing
power for 200-MeV protons incident on $^{40}$Ca.  Figure~\ref{one}
shows elastic scattering while Fig.~\ref{two} presents
inelastic scattering to the $3^-$ state at 3.736~MeV.  The data
are from Ref.\ \cite{HS}.  The solid
curves are based on the chiral NN potential while the long-dashed
curves make use of the conventional OBE potential.  The agreement
between these two is good.  Both calculations include the BHF medium
effects described above.  To demonstrate
the size of these medium corrections, the short-dashed curves are
free-space predictions from the CD-Bonn potential (the chiral force shows
a similar change when medium effects are removed).
Such a change is much larger than the differences between the
chiral and OBE potentials.

Previous studies \cite{S1} have shown that a relativistic treatment
of the medium improves the analyzing power for elastic scattering and
inelastic scattering to natural-parity states, which is where medium
effects are most clearly seen through the central and spin-orbit
components of the interaction. Thus,
 the differences between the data and the
two BHF curves may indicate problems with the treatment of the
medium rather than with the basic interaction.  Compared to
these theoretical uncertainties, the difference between the chiral
and OBE calculations is small, and we consider the chiral
calculations shown here for proton-induced reactions to be
very satisfactory.
Since elastic scattering and the excitation of the $3^-$
state are both
natural parity transitions, this good agreement
applies primarily to the
central and spin-orbit components of the effective NN interaction.
These are the largest parts of the isoscalar interaction.

An evaluation of the isovector part of the chiral potential is
best made by a comparison to a high-spin, unnatural-parity (p,p$'$)
transition.  A set of data that includes polarization transfer
can be used to judge the correct size of each of the spin-dependent
terms in the NN amplitude \cite{Moss,Blez}.  For the highest spins,
the transition is reduced to a simple particle-hole
excitation in which the transferred $J$ is the sum of the
individual contribution: $J=j_{\rm part}+j_{\rm hole}$.  Such
transitions take place in the nuclear surface where the density
is small and usually show minimal effects from the density
dependence of the effective interaction.

\begin{figure}[t]
\vspace{-6.5cm}
\hspace{-1.5cm}
\epsfig{figure=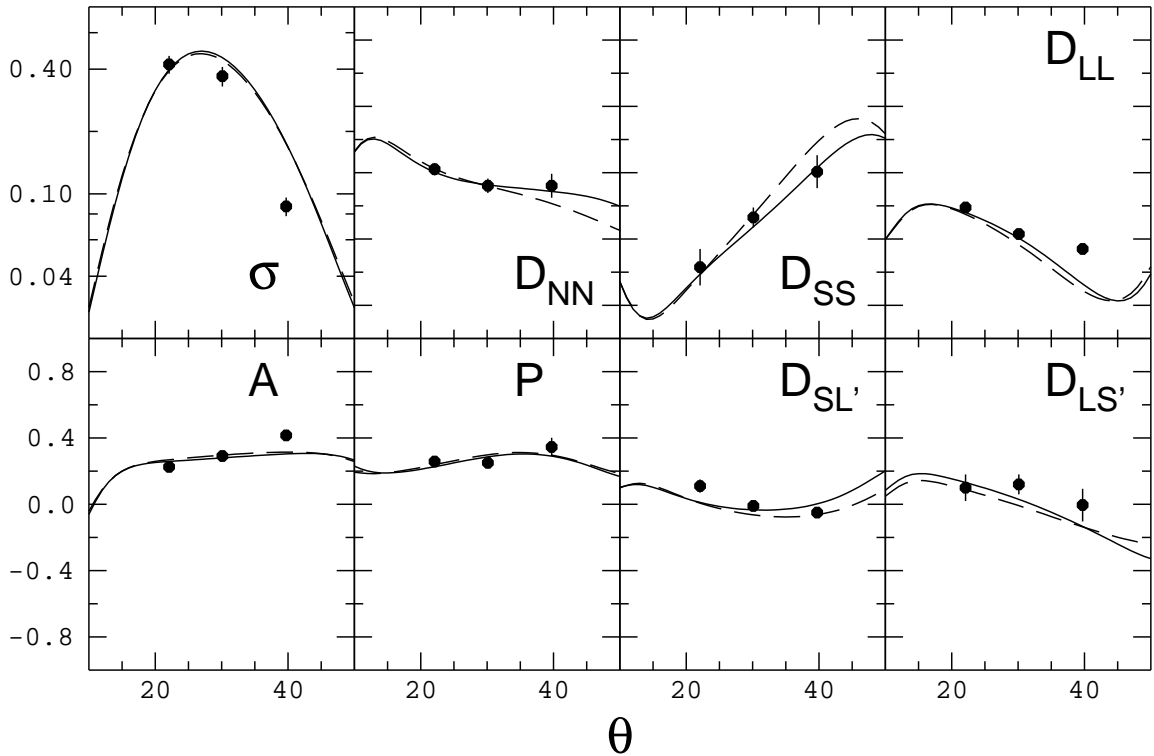,width=15cm}
\vspace{-2cm}
\caption{Measurements of the cross section and polarization
observables for the transition to the $4^-$, T=1 state at 18.98~MeV
in $^{16}$O from Refs.~[15,16].
The solid (dashed) curves are based on the chiral
potential of Ref.~[1] (the CD-Bonn potential of Ref.~[9]).}
\label{three}
\end{figure}

Figure~\ref{three} presents the chiral and
CD-Bonn calculations (solid
and dashed curves, respectively) for the $4^-$, T=1 transition in
$^{16}$O to the state at 18.98~MeV.  BHF density dependence is
included.
The measurements are taken from Refs.~\cite{Opper,Olmer}.
The agreement between
the two potentials, and with the measurements, is excellent.
Interestingly, the chiral potential shows better agreement
with the data for
$D_{NN}$ and $D_{SS}$ than does the original CD-Bonn potential.
This improvement comes from a small reduction in the
spin-longitudinal amplitude (associated with the $\sigma_{1q}
\sigma_{2q}$ tensor operator).

\begin{figure}
\vspace{-6.5cm}
\hspace{-1.5cm}
\epsfig{figure=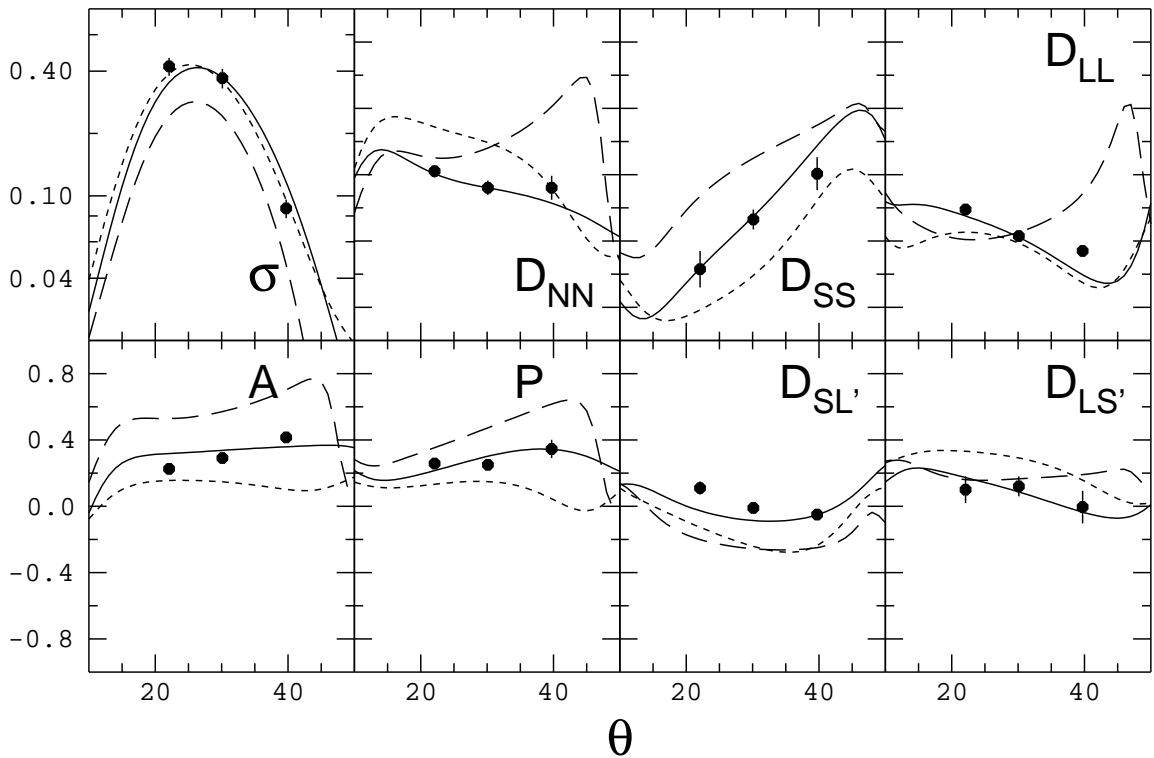,width=15cm}
\vspace{-2cm}
\caption{The measurements are described in Fig.~3.  The solid curves
are based on the chiral potential of Ref.~[1].  The long- and
short-dashed curves are based on the NLO and NNLO potentials of
Ref.~[7].}
\label{four}
\end{figure}

In EM \cite{EM}, the reproduction of the NN phase shifts up to
300~MeV was compared to the predictions from the second order [or
next-to-leading order (NLO)] and the third order [or
next-to-next-to-leading order (NNLO)] potentials of Ref.~\cite{Eb1}.
The second order interaction from that work has been used recently
as the basis for Faddeev calculations of three-body observables
\cite{Eb2}.  Some success was found for energies near and below
10~MeV.  However, at 200~MeV the phase shift predictions
diverge \cite{EM}.  So it is important to see at what level
these differences also appear in (p,p$'$) reactions.  In
Fig.~\ref{four}, we again show the polarization measurements
for the $4^-$, T=1 state in $^{16}$O.  The solid curves are
based on the EM chiral model.  In this case, we are showing free-space
predictions, since we are mainly interested in identifying
baseline differences among the various forces. Furthermore, medium
effects for this transition are small (compare the solid curves
in Fig.~\ref{three} and Fig.~\ref{four}), and thus would
not alter the picture in a significant way.
The long-dashed and
short-dashed curves in Fig.~\ref{four} show the NLO and NNLO interactions
of Ref.~\cite{Eb1}, respectively. The NNLO contains
pion-exchange contributions to the same order as in EM.  The main
difference is that EM have included contact terms to fourth
order and increased the number of momentum cutoff parameters.
While this increases the number of free parameters to be
determined from the NN phase shifts, it also provides the
flexibility necessary for an accurate fit at higher energies.
The NLO and NNLO curves shown in Fig.~\ref{four} both differ
dramatically from the measurements and even at NNLO do not
appear to be converging.  These differences exceed by an order
of magnitude those shown in Fig.~\ref{three}.  This sets a scale
for how much better the phase shift reproduction must be before
it makes sense to compare
these chiral potential predictions with nuclear
reaction measurements at the level of conventional models.
The chiral potential of EM meets this standard.

The $\chi$PT model of EM gives rise to a number of solutions
that differ in some of their short-range characteristics.  This is
illustrated by excellent agreement with the long-range
properties of the deuteron (binding energy, quadrupole moment,
asymptotic S- and D-states, and the mean radius) while allowing
the D-state probability to vary by a factor of two.  These
solutions provide comparable fits to NN phase shifts.  Thus the handling
of the short-range part via contact terms brings about a larger degree
of flexibility as compared to the usual meson exchange picture
(where, for instance, the strength of the tensor force as measured
from the deuteron D-state probability is much more tightly constrained).
Because
of the restrictions imposed by the $\chi$PT expansion on the
typical momenta involved in NN scattering or a nuclear reaction,
it may not be possible to explore and control these ambiguities by going to
higher energies.  Instead, we may need to examine other nuclear
reactions in situations that emphasize the upper end of the
allowed momentum range, such as one finds in large angle
scattering or where the only contributing amplitudes come from
nucleon exchange. Such an investigation is beyond the scope of this paper.

The test calculations shown here demonstrate that
$\chi$PT models of the NN interaction can be made with sufficient
accuracy to be used in calculations of nucleon-induced reactions
on nuclei, at least up to 200~MeV.  For this, a high precision
reproduction of the NN scattering phase shifts is an essential
requirement.  This means that models intended for wide application
must contain a sufficient amount of flexibility to make such
a high precision reproduction possible.  It will be the object of a future
work to explore further the predictive power of high-precision
potentials based on $\chi$PT.

\acknowledgements

The authors acknowledge financial support from the U.S. \
Department of Energy under grant No.\ DE-FG03-00ER41148 (F.S. and
D.A.) and from the National Science Foundation under grant
NSF-PHY-9602872 (E.S.).

\end{document}